\documentclass[final,3p,times]{elsarticle}
\usepackage{epsfig,natbib}
\usepackage{amssymb, amsmath}
\usepackage{lineno}
\usepackage{xcolor, ulem}

\journal{Physica D}
\begin{document}
\begin{frontmatter}

\title{Asymmetric dual cascade in gravitational wave turbulence}
\author[lpp]{Benoît Gay}
\author[lpp]{Sébastien Galtier}
\affiliation[lpp]{
    organization={Laboratoire de Physique des Plasmas, École Polytechnique, Université Paris-Saclay, CNRS},
    addressline={route de Saclay},
    postcode={91128},
    city={Palaiseau},
    country={France}
}

\begin{abstract}
    We numerically simulate, in both the forced and decay regimes, a fourth-order nonlinear diffusion equation derived from the kinetic equation of gravitational wave turbulence in the limit of strongly local quartic interactions. When a forcing is applied to an intermediate wavenumber $k_i$, we observe a dual cascade of energy and wave action. In the stationary state, the associated flux ratio is proportional to $k_i$, and the Kolmogorov-Zakharov spectra are recovered. In decaying turbulence, the study reveals that the wave action spectrum can extend to wavenumbers greater than the initial excitation $k_i$ with constant negative flux, while the energy flux is positive with a power law dependence in $k$. This leads to an unexpected result: a single inertial range with a Kolmogorov-Zakharov wave action spectrum extending progressively to wavenumbers larger than $k_i$. We also observe a wave action decay in time in $t^{-1/3}$ while the front of the energy spectrum progresses according to a $t^{1/3}$ law. These properties can be understood with simple theoretical arguments. 
\end{abstract}



\end{frontmatter}

\section{Introduction}
\label{sec:introduction}

The essence of turbulence is the existence of a scale-by-scale transport of a quantity, like energy, from an initial wavenumber $k_i$ to generally a final wavenumber $k_\infty \gg k_i$, beyond which viscous effects become dominant. We call this phenomenon a cascade, and this range of scales the inertial range \cite{GaltierCUP2023}. In classical three-dimensional hydrodynamic (strong or eddy) turbulence, a direct cascade of energy from large to small scales is observed experimentally and numerically \cite{Grant1962,Kaneda2006}. This phenomenon is also found in many systems \cite{Alexakis2018}. However, nature is sometimes more complex, and a dual cascade can appear. For example, in two-dimensional eddy turbulence, enstrophy undergoes a direct cascade, while energy cascades inversely \cite{Paret1997, Chertkov2007}. A dual cascade is also found in other systems, such as three-dimensional strong magnetohydrodynamic turbulence, with a direct cascade of energy and an inverse cascade of magnetic helicity \cite{Pouquet1976}. 

Although the concept of a dual cascade in eddy turbulence has been recognized since the pioneering contributions of Kraichnan \cite{Kraichnan1967, Kaner1970}, formulating a comprehensive theory to describe this phenomenon remains a significant challenge. In contrast, wave turbulence escapes these difficulties. This regime focuses on the long-term statistical dynamics of a set of weak, random waves interacting through non-linear processes. 
Initially developed to study surface waves \cite{Hasselmann1962, Zakharov1967}, wave turbulence has been extended to many other areas of physics, such as astrophysical plasmas, non-linear optics, acoustics or quantum mechanics, to name but a few \cite{Zakharov1970, Dyachenko1992, Galtier2000, Galtier2003, During2006, Nazarenko2006BEC, Mordant2008, Nazarenko2011, Clark2014, Yarom2014, Monsalve2020, Ricard2021, Falcon2022, Dematteis2023, Galtier2023, David2024, Kochurin2024, Labarre2024, Shavit2024}. 

The first main result of the wave turbulence theory is the derivation of a kinetic equation that describes the temporal evolution of the energy spectrum (or any other type of conserved quantity). This equation can be obtained asymptotically using a multiple time scale technique \cite{Benney1966, Newell1968, gay_2024}; today, wave turbulence theory has even reached the level of a theorem \cite{Deng2023}. When four-wave interactions dominate, the wave action can be a conserved quantity if certain symmetries are satisfied by the kinetic equation. In this case, we generally find an inverse cascade of wave action. We are led to a dual cascade as the energy undergoes a direct cascade. Unlike strong turbulence, wave turbulence can be analyzed in great detail: the exact stationary solutions known as Kolmogorov-Zakharov spectra can be derived analytically, the direction of the cascades can be proved, and the Kolmogorov constants can be derived. 

Recently, another example has appeared in wave turbulence that concerns cosmology. This is gravitational wave turbulence, in which the main contribution comes from four-wave interactions with a kinetic equation with sufficient symmetry to conserve both energy and wave action \cite{galtier_2017}. Pioneering work in this field has led to the conclusion that the inverse cascade is explosive due to the finite capacity of the system for wave action, while it is infinite for energy. In this analytical study, where the gravitational waves propagate in flat space-time, the exact solutions of the kinetic equation known as Kolmogorov-Zakharov spectra were also derived. 

As the direct numerical simulation of such a system is somewhat cumbersome \cite{Galtier2021}, following previous studies for modeling eddy or wave turbulence \cite{Leith1967, Zakharov1999, Connaughton2004, Nazarenko2006, Galtier2020}, a second-order non-linear diffusion model was proposed to mimic the turbulence of weak gravitational waves in the limit of strongly local quartic interactions. Simulations of this diffusion model revealed that, during the explosive phase, the non-stationary wave action spectrum has an anomalous scaling that is understood as a self-similar solution of the second kind \cite{galtier_2019}. More recently, a fourth-order non-linear diffusion model has been proposed to potentially reproduce both the inverse cascade of wave action and the direct cascade of energy \cite{thalabard_2021}. However, the numerical study has not addressed this issue. The aim of this article is precisely to study this dual cascade under different conditions. 

The paper is presented as follows. In section \ref{sec:nonlinear_diffusion_equation}, we introduce the fourth-order nonlinear diffusion equation derived from the kinetic equation. The detail of the derivation is given in \ref{appendix:derivation}. We present the numerical setup in section \ref{sec:numerical_setup} and the results of the numerical simulations in section \ref{sec:numerical_simulations}. A conclusion closes the paper in section \ref{sec:conclusion}. 

\section{Nonlinear diffusion equation}
\label{sec:nonlinear_diffusion_equation}

The kinetic equation for gravitational wave turbulence implies four-wave interactions. It has been shown in \cite{galtier_2017, gay_2024} that in the isotropic case, this equation takes the following form (see \ref{appendix:derivation} for more details):
\begin{equation}
	\frac{\partial N_k}{\partial t}=  
    36 \pi \int_{(\mathbb{R}^{+})^3} \mathcal{T}^{k k_1}_{k_2 k_3}  \left( \frac{k}{N_k} + \frac{k_1}{N_{k_1}} - \frac{k_2}{N_{k_2}} - \frac{k_3}{N_{k_3}} \right) N_k N_{k_1} N_{k_2} N_{k_3} \delta^{01}_{23}(k)~\mathrm{d}k_1 \mathrm{d}k_2 \mathrm{d}k_3,
	\label{eq:isotropic_kinetic_equation}
\end{equation}
where $N_k=N(k)$ stands for the one-dimensional (1D) wave action spectrum, $\mathcal{T}^{k k_1}_{k_2 k_3}$ is the interaction coefficient with a degree of homogeneity equals to $-2$ (see \ref{appendix:derivation}), and $\delta^{01}_{23}(k) = \delta(k + k_1 - k_2 - k_3)$ is a Dirac delta function. Equation (\ref{eq:isotropic_kinetic_equation}) conserves the total wave action $N^\mathrm{tot} = \int_{\mathbb{R}^+} N_k~\mathrm{d}k$ and the total energy $E^\mathrm{tot} = \int_{\mathbb{R}^+} k N_k~\mathrm{d}k$. 

Assuming that the 1D spectrum of wave action is a power law in $k$ such that $N_k = C_N k^x$ (with $C_N \ge 0$), two trivial solutions emerge: $x=0$ and $x=1$. The right-hand side of equation \eqref{eq:isotropic_kinetic_equation} vanishes for these values: they correspond to the thermodynamic equilibrium (null flux) in which there is equipartition of wave action (for $x=0$) and energy (for $x=1$). However, these are not the only available solutions. Indeed, the Zakharov transformation \cite{ZLF92} gives two other solutions, $x = - 2/3$ and $x=-1$, which are associated with non-zero constant fluxes. 
The sign of the associated fluxes can be used to determine the direction of the cascades. One can show that for $x = - 2/3$, the wave action flux $\mathcal{Q}$ is negative, and the associated cascade is inverse. On the other hand, for $x = - 1$, the energy flux $\mathcal{P}$ is positive, and the associated cascade is direct.

In the limit of strongly local quartic interactions, one can show that the integro-differential equation (\ref{eq:isotropic_kinetic_equation}) can be reduced to a fourth-order nonlinear diffusion equation. (For three-wave interactions, the strongly local limit leads to a second-order nonlinear diffusion equation \cite{GaltierCUP2023}.) This derivation is presented in \ref{appendix:derivation}. The diffusion equation for 2D gravitational wave turbulence takes the following form (the 3D case is discussed in \cite{thalabard_2021})
\begin{equation}
    \frac{\partial N_k}{\partial t} = \frac{\partial^2}{\partial k^2} \left[k^4 N_k^4 \frac{\partial^2}{\partial k^2}\left(\frac{k}{N_k}\right) \right].
    \label{eq:nonlinear_diffusion_model}
\end{equation}
Note that the time is normalized to the cascade time for four-wave interactions. 

To assess the validity of the model, we have to recover the exact solutions (thermodynamic and Kolmogorov-Zakharov spectra) for wave action $N_k$ and energy $E_k=kN_k$. 
With $N_k = C_N k^x$, the wave action flux is written as (with $\partial N_k / \partial t = - \partial \mathcal{Q} / \partial k$)
\begin{equation}
    \mathcal{Q}(k) = - C_N^3 x (x - 1) (3 + 3x) k^{2 + 3x}.
\end{equation}
It follows that $\mathcal{Q}$ vanishes when $x = 1$ or $x = 0$, reproducing the behavior of thermodynamic solutions. Moreover, for $x = -2/3$, the flux is constant and equal to $-10 C_N^3 / 9$, which is negative. This result confirms the expected behavior of an inverse cascade for the wave action. Additionally, the Kolmogorov constant can be derived analytically for this model and is given by: $C_K^N = N_k \left( - \mathcal{Q} \right)^{-1/3} k^{2/3} = \left( 9 / 10 \right)^{1/3}$.

Similarly, the energy flux is written as (with $\partial E_k / \partial t = - \partial \mathcal{P} / \partial k$):
\begin{equation}
    \mathcal{P}(k) = - C_N^3 x (x-1) (2+3x) k^{3+3x}.  
\end{equation}
It is straightforward to see that $\mathcal{P}$ vanishes when $x=1$ or $x=0$. For $x=-1$, the energy flux is constant and equal to $2 C_N^3$, which is positive, meaning that the energy cascade is direct. In this case, the Kolmogorov constant is given by: $C_K^E = N_k \mathcal{P}^{-1/3} k = \left( 2 \right)^{-1/3}$.

\section{Numerical setup}
\label{sec:numerical_setup}

This study focuses on the numerical simulation of the previously described system, incorporating dissipation and forcing as governed by the following equation:
\begin{equation}
    \frac{\partial N_k}{\partial t} = \frac{\partial^2}{\partial k^2} \left[k^4 N_k^4 \frac{\partial^2}{\partial k^2}\left(\frac{k}{N_k}\right) \right] - \mathcal{D}_k N_k + \mathcal{F}_k(N_k, t),
\label{eq:simulation_equation}
\end{equation}
where $\mathcal{F}_k(N_k, t)$ is a time-dependent forcing term, and 
$\mathcal{D}_k$ stands for a dissipative term which includes hypo- and hyper-viscosities, denoted $\nu_0$ and $\nu_\infty$ respectively, such that
\begin{equation}
    \mathcal{D}_k \approx \nu_0 k^{-4} + \nu_\infty k^4.
\label{eq:dissipation}
\end{equation}
In practice, instead of using $\nu_0$ and $\nu_\infty$, we define two wavenumbers $k_l$ and $k_h$ at which dissipation is dominant (see \ref{appendix:numerics}). 

The numerical resolution of this equation is performed on a logarithmically refined grid in wavenumber space, enabling the resolution of an extended inertial range. Throughout the run, we follow the evolution of an initial Gaussian distribution of wave action centered around a wavenumber $k_i$ with a standard deviation $\sigma_i$. For such a distribution, the integrals $N^\mathrm{tot} = \int N_k \mathrm{d}k$ and $E^\mathrm{tot} = \int k N_k \mathrm{d}k$ converge, meaning it has a finite wave action and finite energy.

As noted by \cite{thalabard_2021}, this kind of numerical simulation requires careful attention to the propagation of sharp fronts having discontinuous gradients. These instabilities can cause numerical blow-ups when employing conventional differentiation schemes. Thus, we adopt the same numerical strategy to address this challenge, implementing a smooth, noise-robust differentiator \cite{holoborodko_2008}. Time integration uses a two-step Adams-Bashforth method with an adaptive time step. At each iteration, the new time step is determined using a Courant–Friedrichs–Lewy condition \cite{courant_1928}.

When a forcing term is applied at each iteration, its amplitude is chosen to compensate exactly for the dissipated wave action, keeping the total wave action $N^\mathrm{tot}$ constant. Further details, such as the exact expressions of $\mathcal{D}_k$ and $\mathcal{F}_k(N_k, t)$ or their implementation are given in \ref{appendix:numerics}.

The main parameters of the runs are summarized in Table \ref{tab:simulations_parameters}.
\begin{table}[hbtp]
    \centering
    \begin{tabular}{ccccccccccc}
        \hline \hline
        name & $k_\mathrm{min}$ & $k_\mathrm{max}$ & $k_i$ & $\sigma_0$ & $N^\mathrm{tot}$ & $k_l$ & $k_h$ & \texttt{forcing} & \texttt{niter} \\ 
        \hline
        \texttt{run1} & $5.96 \cdot 10^{-8}$ & $1.68 \cdot 10^7$ & $1.0$ & $0.1$ & $10^{20}$ & $10^{-6}$ & $10^{6}$ & yes & $1 \cdot 10^9$ \\
        \texttt{run2} & $5.96 \cdot 10^{-8}$ & $1.68 \cdot 10^7$ & $1.0$ & $0.1$ & $10^{20}$ & $10^{-6}$ & $10^{6}$ & no  & $1 \cdot 10^9$ \\
        \hline \hline
    \end{tabular}
    \caption{Summary of the different parameters used in the simulations, $k_\mathrm{min}$ and $k_\mathrm{max}$ are respectively the lowest and the highest resolved wavenumber in the grid, $k_i$ is the injection wavenumber, $\sigma_i$ is the standard deviation of the initial condition, $N^\mathrm{tot}$ is the initial total amount of wave action, $k_l$ is the wavenumber below which hypo-dissipation starts, $k_h$ is the wavenumber above which hyper-dissipation starts, \texttt{forcing} determines whether a forcing is applied or not, and \texttt{niter} is the total number of iterations.}
    \label{tab:simulations_parameters}
\end{table}



%

\section{Numerical simulations}
\label{sec:numerical_simulations}

\subsection{Forced turbulence}

We first present the numerical simulation \texttt{run1} with a non-zero forcing term (see equation (\ref{eq:simulation_equation})). This is the situation implicitly assumed when the Kolmogorov-Zakharov spectra are derived analytically since a stationary state can only be reached in the presence of a forcing that balances (in the statistical sense) the large-scale or small-scale dissipation. 

\begin{figure}[!ht]
    \centering
    \includegraphics[width=0.8\linewidth]{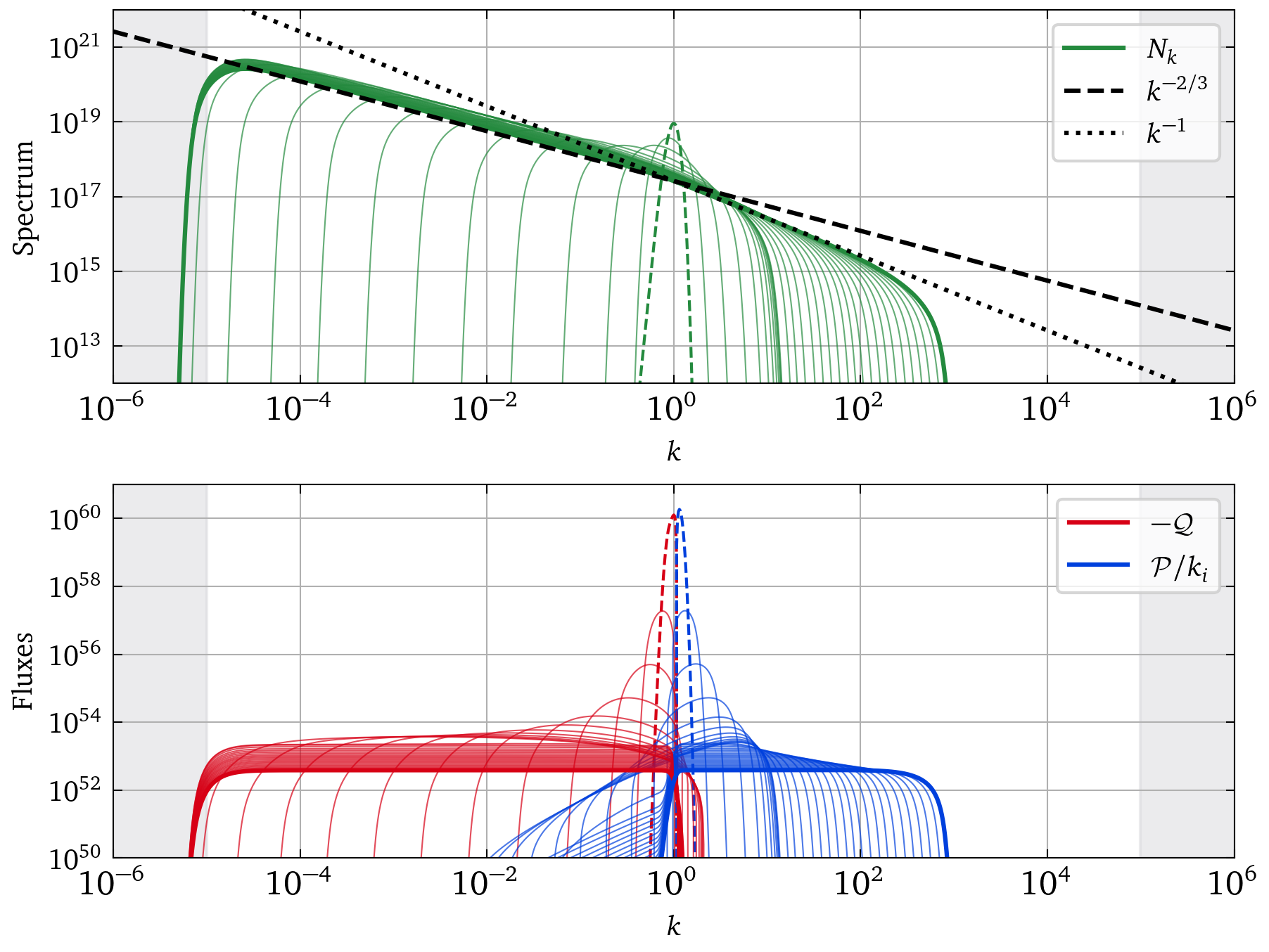}
    \caption{Top: Temporal evolution of the wave action spectrum with a forcing applied at $k_i=1$. Bottom: temporal evolution of the wave action and energy fluxes (at the same times). The grey areas correspond to regions where hypo-dissipation (left) and hyper-dissipation (right) dominate the dynamics (see \ref{appendix:numerics}).}
    \label{fig:forced_spectra}
\end{figure}

Figure \ref{fig:forced_spectra} (top) shows the temporal evolution of the wave action spectrum with a forcing applied at $k_i=1$. As expected, the spectrum propagates faster at low $k$ than at high $k$. This is because the inverse cascade of wave action is explosive as the system has a finite capacity to accumulate wave action at $k<k_i$ (with $N_k \sim k^{-2/3}$, the total wave action at $k<k_i$ is finite). In contrast, the system has an infinite capacity to accumulate energy at $k>k_i$ (with $E_k \sim k^0$, this gives an infinite total energy at $k>k_i$ ), so the direct cascade is asymptotically slow. We also see that the Kolmogorov-Zakharov spectra are found (with the relation $N_k = E_k /k$, the expected wave action spectrum is $N_k \sim k^{-1}$ at large $k$). Note that the hyper-viscous scale is still not reached by the end of the simulation, whereas the hypo-viscous scale is. 
Figure \ref{fig:forced_spectra} (bottom) shows the corresponding wave action and energy fluxes. We observe a rapid drop in flux amplitude during the propagation towards low $k$ of the spectral front $k_0(t)$ of wave action. This evolution is the consequence of the conservation of the total wave action over time (not shown) imposed by the forcing, which leads to a decrease in the wave action flux. Then, constant fluxes in $k$ are found with a slow development of the inertial range to high $k$. The two flux amplitudes slowly decrease at the same rate. This last property can again be understood as a consequence of the total wave action conservation (over time) coupled with the propagation towards high $k$ of the spectral front $k_\infty(t)$ of energy. This leads to a decrease over time of the energy flux, and therefore in the wave action flux, as the two are linked by the relation $\mathcal{P}=-k_i \mathcal{Q}$ (the forcing injects lower flux in order to conserve the total wave action). 
We can notice that simulations (not shown) with forcing applied at $k_i \ll 1$ or $k_i \gg 1$ show the same qualitative behavior, which can thus be considered universal in stationary turbulence. 

\begin{figure}[!ht]
    \centering
    \includegraphics[width=0.45\linewidth, trim=0 0 290 0, clip]{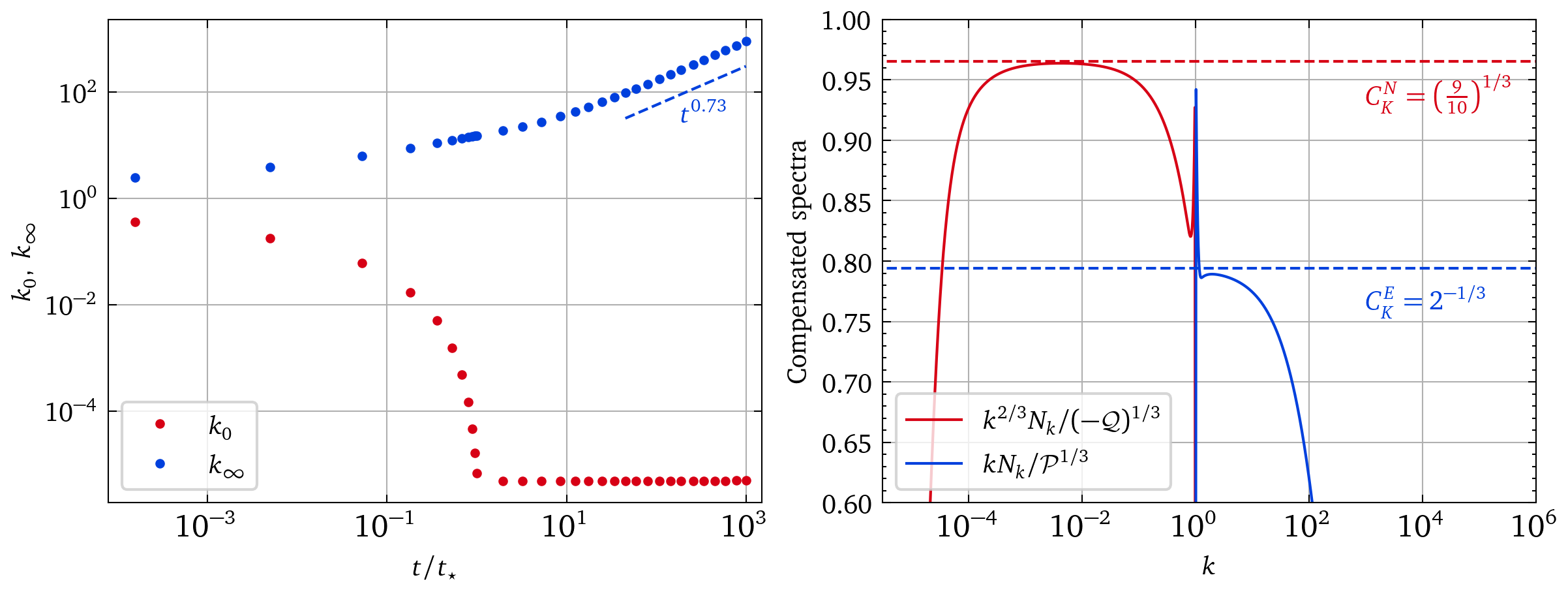}
    \hfill
    \includegraphics[width=0.45\linewidth, trim=285 0 0 0, clip]{pictures/forced_0_fronts.png}
    \caption{Left: Temporal evolution of the spectral front of wave action $k_0$ (red) and energy $k_\infty$ (blue). The time is normalized by $t_* \approx 1.38 \cdot 10^{-36}$ (see text). Note that the dots are plotted at the same times as the spectra in Figure \ref{fig:forced_spectra}. 
    Right: Estimation of the Kolmogorov constants derived from the wave action and energy spectra. }
    \label{fig:forced_fronts}
\end{figure}

Figure \ref{fig:forced_fronts} (left) shows the propagation of the spectral fronts $k_0(t)$ and $k_\infty(t)$. 
Numerically, the positions of the fronts are determined by following the propagation of wave action spectrum for the value $N_k=10^{10}$.
This value is chosen to be sufficiently high to eliminate the slight amplitude fluctuation and sufficiently low to cross the spectra at each iteration.
As expected, we first observe a sharp decrease in $k_0(t)$, which means that in principle, $k_0(t)$ can reach the mode $k = 0$ in a finite time $t_*$ \citep{galtier_2019}. This time, $t_*$ is used to normalize the plot. The hypo-viscosity scale is then reached, and $k_0(t)$ no longer changes. Moreover, we can see that $k_\infty(t)$ increases slowly with a power close to $t^{0.72}$, which means in principle (in the inviscid limit) that it takes an infinite time to get $k_\infty(t)=+\infty$. 
%
Figure \ref{fig:forced_fronts} (right) shows a measure of the Kolmogorov constants compared with their theoretical predictions. The Kolmogorov constant estimated numerically for the wave action is $C_K^N \simeq 0.965$, which fits well the theoretical value $\left( \frac{9}{10}\right)^{1/3}$. 
The Kolmogorov constant found numerically for the energy, $C_K^E \simeq 0.790$, is slightly lower than the theoretical value $2^{-1/3} \approx 0.794$. 
This discrepancy is probably due to the narrow inertial range for the direct cascade.

In conclusion, we have shown that the fourth-order nonlinear diffusion model in the forced case can reproduce the main results expected from the theory, with a dual cascade at constant fluxes. Furthermore, we have shown that the flux of the wave action decreases over time due to physical processes occurring at $k>k_i$, while the forcing maintains a constant total wave action. So, while the model describes super-local quartic exchanges of energy and wave action, we have seen that small-scale dynamics can directly influence large-scale dynamics.

\subsection{Decaying turbulence}

\subsubsection{Numerical simulation}

We now present the numerical simulation \texttt{run2} with a vanishing forcing term (see equation (\ref{eq:simulation_equation})). Hence, only dissipation at low- and high-$k$ occurs.

\begin{figure}[!ht]
    \centering
    \includegraphics[width=.8\linewidth]{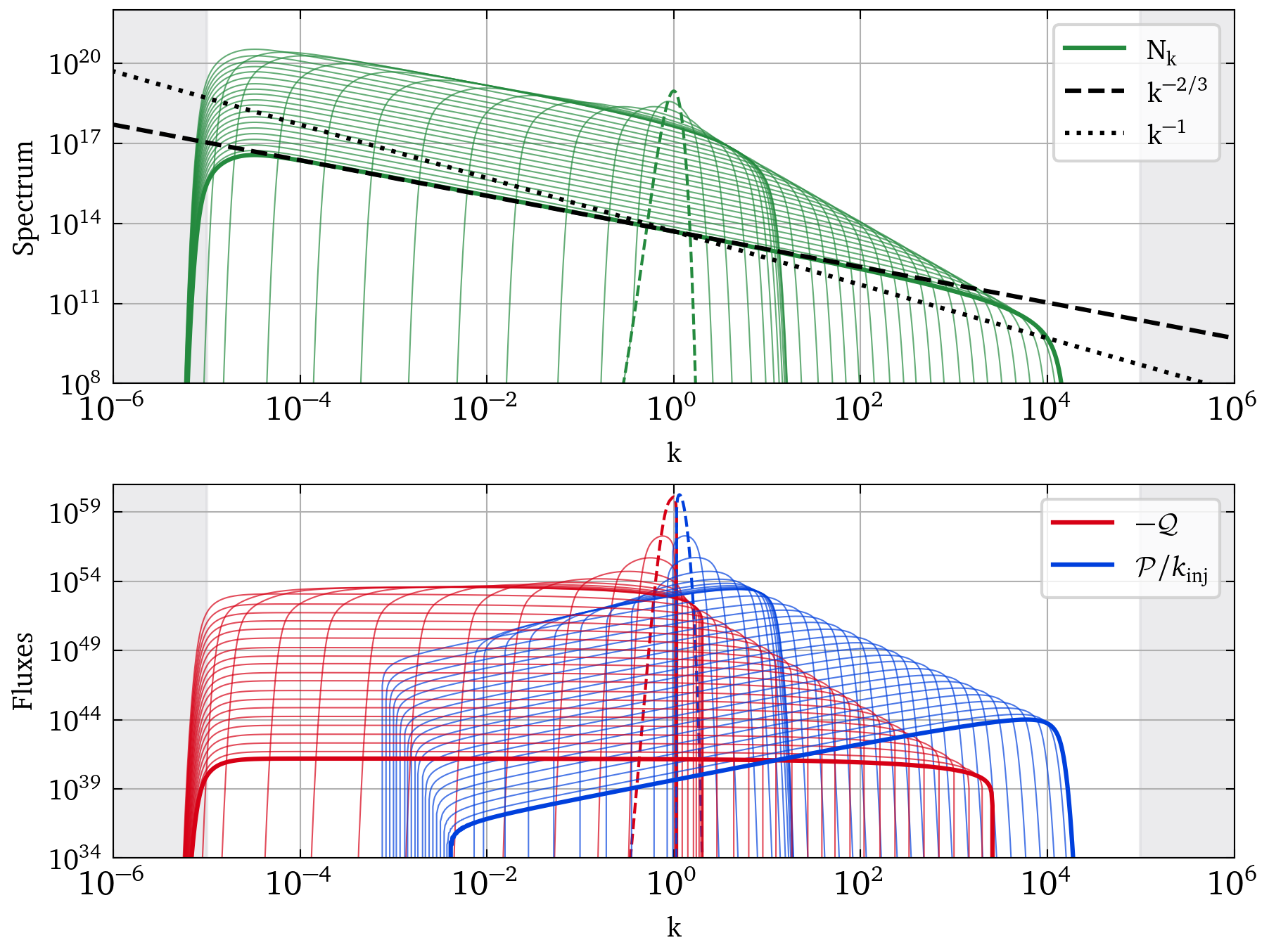}
    \caption{Top: Temporal evolution of the wave action spectrum with an initial excitation applied at $k_i=1$. Bottom: temporal evolution of the wave action and energy fluxes (at the same times). The grey areas correspond to regions where hypo-dissipation (left) and hyper-dissipation (right) dominate the dynamics.}
    \label{fig:free_spectra}
\end{figure}

Figure \ref{fig:free_spectra} shows a simulation of a free decay. Unlike the forced case discussed earlier, the wave action spectrum (top) exhibits a single power law, consistent with the $k^{-2/3}$ Kolmogorov-Zakharov solution. Interestingly, this spectrum propagates in both the low-$k$ and high-$k$ directions, suggesting the presence of both an inverse and a direct cascade. The inverse cascade progresses more rapidly than the direct one, highlighting a notable asymmetry in their dynamics.
To understand this phenomenon, it is helpful to examine the temporal evolution of the fluxes (bottom): while the wave action flux is roughly constant (but not necessarily null for $k>k_i$ as it is in a forced simulation), the energy flux scales in $k$. Injecting this relationship into the Kolmogorov-Zakharov spectrum gives $E_k \sim k^{1/3}$, compatible with the observed $k^{-2/3}$ wave action spectrum. Note that the hyper-viscous scale is still not reached by the end of the simulation, whereas the hypo-viscous scale is. Therefore, this simulation reveals that an overlap between the two inertia ranges is possible when the system is not forced. It also shows that the cascade of energy towards the high $k$ occurs unusually, with the energy flux constrained by a constant wave action flux. This positive flux, which slowly drives the energy spectrum towards $k>k_i$, brings with it wave action. Since wave action has a much shorter time scale than energy, an inverse cascade occurs, leading to the counter-intuitive observation of spectra of an apparent direct cascade of wave action. 
As for the study in the previous section, other simulations (not shown) carried out with $k_i \ll 1$ and $k_i \gg 1$ give the same qualitative behavior. 

\begin{figure}[!ht]
    \centering
    \includegraphics[width=0.45\linewidth, trim=0 0 288 0, clip]{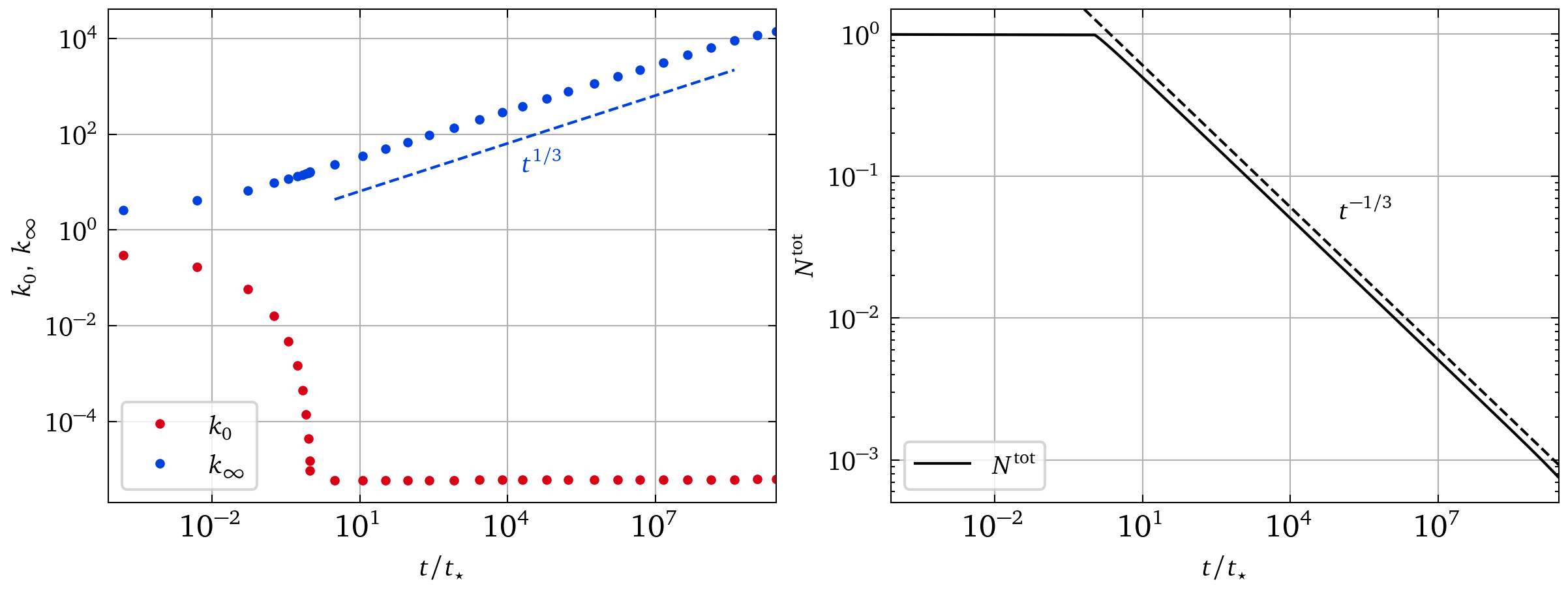}
    \hfill
    \includegraphics[width=0.45\linewidth, trim=288 0 0 0, clip]{pictures/free_0_fronts.png}
    \caption{Left: Temporal evolution of the spectral front of energy $k_\infty$ (blue) and wave action $k_0$ (red). The theoretical prediction in $t^{1/3}$ is plotted in dashed line.
    Right: Temporal decay of the total wave action which follows the theoretical prediction in $t^{-1/3}$ plotted in dashed line. Time is normalized by $t_* \approx 1.38 \cdot 10^{-36}$ (see text).}
    \label{fig:free_fronts}
\end{figure}

Figure \ref{fig:free_fronts} (left) shows the propagation of the spectral fronts $k_0(t)$ and $k_\infty(t)$. Note that the measurements taken on the fronts correspond to the times of the spectra shown in Figure \ref{fig:free_fronts}. 
Once again, they have been numerically computed following the propagation of the wave action spectrum but here for the value $N_k = 10^{8}$.
As in the previous section, we first observe a sharp decrease in $k_0(t)$, meaning that $k_0(t)$ can reach $0$ in a finite time $t_*$ \citep{galtier_2019}, which is used to normalize the time. The hypo-viscosity scale is then reached, and $k_0(t)$ no longer changes. On the other hand, we see that $k_\infty(t)$ increases with a power close to $t^{1/3}$, which means that it takes an infinite time to have $k_\infty(t)=+\infty$. 
Numerically, a linear regression shows an evolution in $t^{0.32}$, slightly less steep than the theoretical prediction. 
We will also show below that this behavior can be predicted theoretically (see below). 
Figure \ref{fig:free_fronts} (right) shows the temporal decay of the total wave action, which follows a $t^{-1/3}$ decay over almost 10 decades. (It is interesting to note that with a direct numerical simulation of gravitational wave turbulence, the law found is close to $t^{-0.25}$ \cite{Galtier2021}). We will show below that this power-law index can be predicted theoretically. 

\subsubsection{Decaying law for wave action}

This section aims to provide a model of decaying turbulence when a dual cascade occurs. To our knowledge, this is the first time such a study has been carried out for wave action, while several energy decay laws have been proposed for strong and weak turbulence \cite{Kolmogorov1941, Galtier1997, Badulin2005}. 

Let us start with region 1 (for $k_0<k<k_i$), where we have an inverse cascade of wave action. We shall assume that the scale $k_0$ below which hypo-dissipation dominates the dynamics is such that $k_0 \ll k_i$. When the spectrum is fully developed, i.e. when the hypo-viscous scales are reached, it follows the stationary Kolmogorov-Zakharov solution:
\begin{equation}
N_k = C_N (-\mathcal{Q})^{1/3} k^{-2/3} ,
\end{equation}
which leads to the total wave action
\begin{equation}
N_1^{tot} = \int_{k_0}^{k_i} C_N (-\mathcal{Q})^{1/3} k^{-2/3} \mathrm{d}k  \simeq C_N (-\mathcal{Q})^{1/3} k_i^{1/3} .
\end{equation}
In region 2 (for $k>k_i$), the system has an infinite capacity to accumulate energy, the direct cascade of energy is slow, and the Kolmogorov-Zakharov solution is:
\begin{equation}
E_k = C_E (\mathcal{P})^{1/3} k^{0} .
\end{equation}
We then deduce the total wave action:
\begin{equation}
N_2^{tot} = \int_{k_i}^{k_\infty} k^{-1} E_k \mathrm{d}k  = \int_{k_i}^{k_\infty} C_E \mathcal{P}^{1/3} k^{-1} \mathrm{d}k  .
\end{equation}

However, the simulation reveals that the energy flux is not constant in region 2 but behaves as $\mathcal{P} = - k \mathcal{Q}$. 
Incorporating this information in the wave action evaluation, we find (with $k_\infty \gg k_i$)
\begin{equation}
N_2^{tot} = \int_{k_i}^{k_\infty} C_E (-\mathcal{Q})^{1/3} k^{-2/3} \mathrm{d}k  \simeq C_E (-\mathcal{Q})^{1/3} k_{\infty}^{1/3} .
\end{equation}
Therefore, the total wave action is
\begin{equation}
N^{tot} = N_1^{tot}+ N_2^{tot} \simeq C_N (-\mathcal{Q})^{1/3} k_i^{1/3} + C_E (-\mathcal{Q})^{1/3} k_{\infty}^{1/3} \simeq C_E (-\mathcal{Q})^{1/3} k_{\infty}^{1/3} .
\end{equation}
On the other hand, the inverse cascade being explosive and the direct cascade slow, we can assume that, in the first phase, the wave action decay happens without energy decay due to viscous effects (in the simulation, the hyper-viscous scale is not reached). 
Under these conditions, the total energy in region 2 is conserved and expressed by:
\begin{equation}
E_2^{tot} = \int_{k_i}^{k_\infty} E_k \mathrm{d}k \simeq C_E (-\mathcal{Q})^{1/3} k_{\infty}^{4/3}.
\label{eq:ener}
\end{equation}
Using expression (\ref{eq:ener}), we obtain:
\begin{equation}
N^{tot} \simeq C_E^{3/4} (-\mathcal{Q})^{1/4} (E_2^{tot})^{1/4} .
\label{eq:ntot}
\end{equation}
We now introduce the definition of wave action decay, expressed as:
\begin{equation}
\frac{d N^{tot}}{dt} = - \int_{k_0}^{k_\infty} \frac{\partial \mathcal{Q}}{\partial k} \mathrm{d}k = \mathcal{Q}(k_0) = \mathcal{Q}.
\end{equation}
Here, we assume, based on numerical observations, that the wave action flux remains nearly constant but strongly decreases before reaching the right front $k_\infty$.

Using expression (\ref{eq:ntot}), we obtain
\begin{equation}
\frac{1}{4}C_E^{3/4} (E_2^{tot})^{1/4} (-\mathcal{Q})^{-3/4} \frac{d  (-\mathcal{Q})}{dt} = \mathcal{Q} ,
\end{equation}
which leads to the following estimate
\begin{equation}
\vert \mathcal{Q} \vert (t) \sim t^{-4/3} ,
\label{eq:qt}
\end{equation}
and finally
\begin{equation}
N^{tot}(t) \sim t^{-1/3} .
\end{equation}
This prediction is in excellent agreement with the numerical simulation (see Figure \ref{fig:free_fronts}). Injecting expression (\ref{eq:qt}) into (\ref{eq:ener}) leads to the following propagation law for the spectral front of energy
\begin{equation}
k_{\infty}(t) \sim t^{1/3} .
\end{equation}
This temporal evolution is also very well observed in the simulation. 
\section{Conclusion}
\label{sec:conclusion}


We have shown that within the framework of a nonlinear fourth-order diffusion model, we can reproduce the expected properties of wave turbulence in the stationary regime. To our knowledge, this is the first attempt to study the dynamics of a dual cascade with such a diffusion model. Furthermore, we showed that in the decaying case, an unexpected behavior emerges with a wave action spectrum that tends to develop not only at $k$ smaller than the injection wavenumber but also at higher values. This property is thought to be due to the characteristic time scale of the inverse cascade, which is much shorter than that of the direct cascade. Consequently, as the direct cascade carries wave action, the latter can have a feedback effect on the dynamics with a tendency to produce an inverse cascade with a constant flux in $k$. This property is then imposed on the energy, which, therefore, has a flux following a scaling proportional to $k$. Finally, we have proposed for the first time a theory that can predict the decay law of wave action when the wave action spectrum is fully developed at low $k$, and the energy spectrum is being developed at high $k$. 
The properties understood within the framework of this fourth-order nonlinear diffusion model are thought to help in better understanding direct numerical simulations of gravitational wave turbulence, which are much more challenging. 

\appendix
\section{From the kinetic equation to the fourth-order diffusion equation}
\label{appendix:derivation}

The starting point of the derivation is the isotropic kinetic equation (\ref{eq:isotropic_kinetic_equation}). The isotropic interaction coefficient has the following properties:
\begin{itemize}
\item it is positive: $\forall k, k_1, k_2, k_3 \in \mathbb{R}^+, \mathcal{T}^{k k_1}_{k_2 k_3} \geq 0$;
\item it is symmetric when permuting the elements: $\mathcal{T}^{k k_1}_{k_2 k_3} = \mathcal{T}^{k_1 k}_{k_2 k_3} = \mathcal{T}^{k k_1}_{k_3 k_2} = \mathcal{T}^{k_2 k_3}_{k k_1}$;
\item it is homogeneous of degree $-2$: $\forall \alpha \in \mathbb{R}^*, \mathcal{T}^{\alpha k \alpha k_1}_{\alpha k_2 \alpha k_3} = \frac{1}{\alpha^2} \mathcal{T}^{k k_1}_{k_2 k_3}$. 
\end{itemize}
Let $f$ be a test function of $C_c^\infty(\mathbb{R}^+)$. The quantity that aims to be computed is
\begin{equation}
    \int_{\mathbb{R}^+} \partial_t N_k f(k)~\mathrm{d}k = 
    \int_{(\mathbb{R}^+)^4} \mathcal{T}^{k k_1}_{k_2 k_3} 
    \left( 
    \frac{k}{N_k} + \frac{k_1}{N_{k_1}} - \frac{k_2}{N_{k_2}} - \frac{k_3}{N_{k_3}} \right) 
    N_k N_{k_1} N_{k_2} N_{k_3} 
    f(k) 
    \delta^{01}_{23}(k)
    ~\mathrm{d}k \mathrm{d}k_1 \mathrm{d}k_2 \mathrm{d}k_3.
    \label{eq:kinetic_isotropic_integrated}
\end{equation}
Taking into account all the symmetries  of the isotropic interaction coefficient $\mathcal{T}^{k k_1}_{k_2 k_3}$, the right-hand side of equation \eqref{eq:kinetic_isotropic_integrated} can be rewritten as
\begin{equation}
    \frac{1}{4} \int_{(\mathbb{R}^+)^4} 
    \mathcal{T}^{k k_1}_{k_2 k_3} 
    \left( 
    \frac{k}{N_k} + \frac{k_1}{N_{k_1}} - \frac{k_2}{N_{k_2}} - \frac{k_3}{N_{k_3}}
    \right) 
    N_k N_{k_1} N_{k_2} N_{k_3}
    \left[f(k) + f(k_1) - f(k_2) - f(k_3) \right] 
    \delta^{01}_{23}(k)
    ~\mathrm{d}k \mathrm{d}k_1 \mathrm{d}k_2 \mathrm{d}k_3.
    \label{eq:kinetic_isotropic_integrated_RHS}
\end{equation}
Now, we assume the locality of the quartic interactions. It means that for each $k$ in $\mathbb{R}^+$, the isotropic interaction coefficient $\mathcal{T}^{k k_1}_{k_2 k_3}$ is equal to zero outside a small region where $k_1$, $k_2$, and $k_3$ are in the interval $\left] k - k \eta, k + k \eta \right[$, where $\eta$ is a fixed positive real, with $\eta \ll 1$. Hence, a good choice of parametrization is
\begin{equation*}
    \forall~ 1 \leq i \leq 3,~k_i = k (1 + \eta_i)
    \quad
    \text{where}
    \quad
    \eta_i \in \left] - \eta, \eta \right[.
\end{equation*}
Consequently, the right-hand side of equation \eqref{eq:kinetic_isotropic_integrated_RHS} can be Taylor-expanded. The different terms become, at first non-vanishing order
\begin{subequations}
\begin{align}
    \mathcal{T}^{k k_1}_{k_2 k_3} &= \frac{1}{k^2} \tilde{\mathcal{T}}^{1 \eta_1}_{\eta_2 \eta_3}; \\
    \left( \frac{k}{N_k} + \frac{k_1}{N_{k_1}} - \frac{k_2}{N_{k_2}} - \frac{k_3}{N_{k_3}} \right) &= \frac{\eta_2^2 + \eta_3^2 - \eta_1^2}{2} k^2 \frac{\partial^2}{\partial k^2}\left(\frac{k}{N_k}\right); \\
    N_k N_{k_1} N_{k_2} N_{k_3} &= N_k^4; \\
    \left[f(k) + f(k_1) - f(k_2) - f(k_3)\right] &= \frac{\eta_1^2 - \eta_2^2 - \eta_3^2}{2} k^2 f''(k); \\
    \delta^{01}_{23}(k) \mathrm{d}k_1 \mathrm{d}k_2 \mathrm{d}k_3 &= k^2 \delta(\eta_1 - \eta_2 - \eta_3) \mathrm{d}\eta_1 \mathrm{d}\eta_2 \mathrm{d}\eta_3,
\end{align}
\end{subequations}
where $\tilde{\mathcal{T}}^{1 \eta_1}_{\eta_2 \eta_3} =  \mathcal{T}^{2 (1+\eta_1)}_{(1+\eta_2) (1+\eta_3)}$ to simplify the notation. 
    
The next step is to introduce these decompositions into the integral \eqref{eq:kinetic_isotropic_integrated_RHS}; it leads to
\begin{equation}
    \frac{1}{16} \int_{]-\eta, \eta[^3} 
    \tilde{\mathcal{T}}^{1 \eta_1}_{\eta_2 \eta_3} \left(\eta_1^2 - \eta_2^2 - \eta_3^2 \right)^2 \delta(\eta_1 - \eta_2 - \eta_3)~\mathrm{d}\eta_1 \mathrm{d}\eta_2 \mathrm{d}\eta_3 
    \times 
    \int_{\mathbb{R}^+} k^4 N_k^4 f''(k)\frac{\partial^2}{\partial k^2}\left(\frac{k}{N_k} \right) ~\mathrm{d}k.
\end{equation}
The first integral is compacted into a constant $C > 0$, and after two integrations by part, the second integral can be written as
\begin{equation}
    \int_{\mathbb{R}^+} f(k) \frac{\partial^2}{\partial k^2} \left[k^4 N_k^4  \frac{\partial^2}{\partial k^2}\left(\frac{k}{N_k}\right) \right] ~\mathrm{d}k.
\end{equation}
Finally, the fourth-order nonlinear diffusion equation is identified by comparing the different sides of equation \eqref{eq:kinetic_isotropic_integrated}. It leads to:
\begin{equation}
    \frac{\partial N_k}{\partial t} = C \frac{\partial^2}{\partial k^2} \left[k^4 N_k^4 \frac{\partial^2}{\partial k^2}\left(\frac{k}{N_k}\right) \right].
    \label{eq:nonlinear_diffusion_equation_weak}
\end{equation}
We eventualy renormalize the time, $t \to C t$, to recover equation (\ref{eq:nonlinear_diffusion_model}).
\section{Numerics}
\label{appendix:numerics}

We gave here further details concerning the implementation of the code used to perform the simulation as presented in section \ref{sec:numerical_setup}.

We solve the equation \eqref{eq:simulation_equation} on a logarithmic-scale grid, $\mathcal{K} = \left\{ k_j = 2^{j / 30}, j \in [-720, 720] \right\}$. The spatial derivative is computed using a smooth noise robust differential differentiator \citep{holoborodko_2008, thalabard_2021}. On each point $k_p$ of the logarithmic-scale grid  $\mathcal{K}$, the first derivative ${\partial N_k}/{\partial k}$ is expressed by:
\begin{equation}
    \frac{\partial N_{k_p}}{\partial k} = \frac{84 \delta_{p,1} + 192 \delta_{p,2} + 162 \delta_{p,3} + 64 \delta_{p,4} + 10 \delta_{p,5}}{512},
\end{equation}
where
\begin{equation}
    \delta_{p,i} = \frac{N_{k_{p+i}} - N_{k_{p-i}}}{k_{p+i} - k_{p-i}}.
\end{equation}

The initial condition is a Gaussian distribution centered around $k_i$ having a standard deviation $\sigma_i$ and having such an amplitude that the total wave action $N^\mathrm{tot}$ is initially fixed. The expression of this initial condition is then given by:
\begin{equation}
    N_k(t=0) = A \times \exp \left[ - \frac{\left(k - k_i \right)^2}{2 \sigma_i^2}\right],
\end{equation}
where the normalization constant $A$ is such that initially: $\sum_{k \in \mathcal{K}} k N_k = N^\mathrm{tot}$.

Furthermore, time integration is performed by using a two-step Adams–Bashforth method having an adaptive time step $\delta t$ computed by using the following Courant–Friedrichs–Lewy condition \cite{courant_1928}:
\begin{equation}
    \delta t = \mu \min_{k \in \mathcal{K}} \frac{N_k + \varepsilon}{| \partial_t N_k + \varepsilon |} ,
\end{equation}
where values of $\varepsilon$ and $\mu$ are $\varepsilon = 10^{-40}$ and $\mu = 5 \times 10^{-3}$ in order to ensure numerical stability and reasonable performance. 

Dissipation and forcing are implemented following Strang's procedure \cite{strang_1968}. At each time step, the solution is advanced by first making the system evolve in the free regime, then applying the dissipation operator, and finally applying the forcing operator. Dissipation is handled using an implicit Euler scheme, which corresponds numerically to multiplying $N_k$ by the constant operator:
\begin{equation}
 \left[ 1 + \left( \frac{k}{k_h} \right)^4\right]^{-1} \left[ 1 + \left( \frac{k}{k_l} \right)^{-4} \right]^{-1}.
\end{equation}
This form ensures dissipation at low $k \leq k_l$ and high $k \geq k_h$ wavenumber. 
Numerical simulations show that hypo-dissipative effects dominate for $k \lesssim 10 k_l$, while hyper-dissipative effects dominate for  $k \gtrsim 0.1 k_h$ (corresponding to a dissipation factor of $\approx 1 - 10^{-4}$). These limits were used in Figures \ref{fig:forced_spectra} and \ref{fig:free_spectra} to define the grey areas.

Forcing is applied after dissipation by adding to the current $N_k(t)$ the initial condition $N_k(t=0)$ scaled by an adaptive constant. This constant is chosen to maintain the total wave action at a constant value.


\bibliographystyle{elsarticle-num-names} 
\providecommand{\noopsort}[1]{}\providecommand{\singleletter}[1]{#1}%

\end{document}